\documentclass{optica-article}

\journal{opticajournal} % for journals or Optica Open

\articletype{Research Article}

\usepackage{lineno}
\begin{document}

\title{Dynamic spectral tailoring of a 10 GHz laser frequency comb for enhanced calibration of astronomical spectrographs}

\author{Pooja Sekhar,\authormark{1,2,*} Connor Fredrick,\authormark{1,2,3} Peter Zhong,\authormark{1} Abijith S. Kowligy,\authormark{4}  Arman Cing\"{o}z,\authormark{4} and Scott A. Diddams\authormark{1,2,\dag}}

\address{\authormark{1}Electrical, Computer and Energy Engineering, University of Colorado Boulder, Colorado 80309, USA\\
\authormark{2}Department of Physics, University of Colorado Boulder, Colorado 80309, USA\\
\authormark{3}National Institute of Standards and Technology, Boulder, Colorado 80305, USA\\
\authormark{4}Vector Atomic, Inc. 6870 Koll Center Pkwy, Pleasanton, CA 94566, USA\\

\email{\authormark{*}Corresponding author: pooja.sekhar@colorado.edu} 
\authormark{$\dag$}email: scott.diddams@colorado.edu\\}

\begin{abstract*} 
Laser frequency combs (LFCs) are an important component of Doppler radial velocity (RV) spectroscopy that pushes fractional precision to the $10^{-10}$ level, as required to identify and characterize Earth-like exoplanets. However, large intensity variations across the LFC spectrum that arise in nonlinear broadening limit the range of comb lines that can be used for optimal wavelength calibration with sufficient signal-to-noise ratio. Furthermore, temporal spectral-intensity fluctuations of the LFC, that are coupled to flux-dependent detector defects, alter the instrumental point spread function (PSF) and result in spurious RV shifts. To address these issues and improve calibration precision, spectral flattening is crucial for LFCs to maintain a constant photon flux per comb mode. In this work, we demonstrate a dynamic spectral shaping setup using a spatial light modulator (SLM) over the wavelength range of 800–1300 nm. The custom shaping compensates for amplitude fluctuations in real time and can also correct for wavelength-dependent spectrograph transmission, achieving a spectral profile that delivers the constant readout necessary for maximizing precision. Importantly, we characterize the out-of-loop properties of the spectral flattener to verify a twofold improvement in spectral stability. This technique, combined with our approach of pumping the waveguide spectral broadener out-of-band at 1550 nm, reduces the required dynamic range. %to approximately 1.3 dB and 2.5 dB respectively for in-loop and out-of loop measurements. (New full BW old case: 2.4 dB (in-loop) and 3.7 dB (out-of-loop))
While this spectral region is tailored for the LFC employed at the Habitable-zone Planet Finder (HPF) spectrograph, the method is broadly applicable to any LFC used for astronomical spectrograph calibration. %Beyond astronomy, our spectral shaping setup, with its broad spectral coverage and high resolution, has potential applications in a wide range of fields including telecommunications, 2D spectroscopy, and holography.

\end{abstract*}

%%%%%%%%%%%%%%%%%%%%%%%%%%  body  %%%%%%%%%%%%%%%%%%%%%%%%%%
\section{Introduction}

The search for Earth-analogs using the radial velocity technique requires precision at the cm/s level, corresponding to fractional Doppler shifts of $10^{-10}$. This precision amounts to a shift in the centroid of stellar absorption lines at the level of 10's of kHz. Functionally, this requires measuring a spatial shift of the dispersed spectrum by about $10^{-5}$ of a pixel in the detector array at the spectrograph image plane. This demanding precision is built upon the long-term frequency stability of a required calibration source, ideally over many years of nightly operation. 

In this regard, laser frequency combs (LFCs) have emerged as ideal calibration sources for astronomical spectrographs since the frequency of all comb teeth can be referenced to a GPS-disciplined atomic clock \cite{braje2008astronomical, murphy2007high, herr2019astrocombs, steinmetz2008laser}. Two critical requirements for ``astrocombs," which are LFCs designed specifically for astronomical spectrographs, include tens of gigahertz (GHz) mode spacing and spectral coverage exceeding 500 nm. The latter requirement of broadband spectral coverage is realized using dispersion-engineered nonlinear media like photonic crystal fibers \cite{probst2020crucial}, highly nonlinear single-mode fibers \cite{ycas2012demonstration, obrzud2018broadband} or waveguides \cite{metcalf2019stellar}. However, the intensity profile of the resulting broadband supercontinuum is not flat and often exhibits wavelength-dependent variations over several orders of magnitude. Additionally, the spectral envelope can fluctuate over time due to changes in the peak power of the pump and the aging of components, particularly the amplifiers and nonlinear media. 

These intensity fluctuations in the comb lines change the continuum level in the detector readout asymmetrically due to overlapping contributions from adjacent lines. This, combined with flux-dependent detector defects such as nonlinear response, brighter-fatter effect, persistence contamination from the previous exposure, and intra- and inter pixel quantum efficiency variations, result in changes to the shape of the instrumental point spread function (PSF) \cite{lage2017measurements, plazas2017nonlinearity, ninan2019impact}. All of these variations can lead to wavelength calibration errors and spurious RV shifts. Additionally, it is essential to maintain a uniform flux distribution while scanning the laser frequency comb (LFC) across the spectrograph, as demonstrated in our previous work \cite{sekhar2024tunable}. Such scanning is crucial for characterizing the impact of intrapixel quantum efficiency variations on the PSF profile, which is key to improving RV precision. Large intensity variations across the spectrum also lead to detector saturation at certain frequencies, while other comb lines may be buried in noise. This results in reduced photoelectron counts and a higher photon noise floor. 

To address the above issues, a dynamic spectral flattener is required for astrocombs. This flattener ensures a uniform photon flux across all comb lines, thereby maximizing the achievable signal-to-noise ratio (SNR) and enhancing the calibration precision of astronomical spectrographs. Moreover, a stable and uniform photon flux of LFC also enables the systematic subtraction of scattered LFC contributions on the stellar spectrum.
\begin{figure}[ht!]
\centering\includegraphics[width=\linewidth]{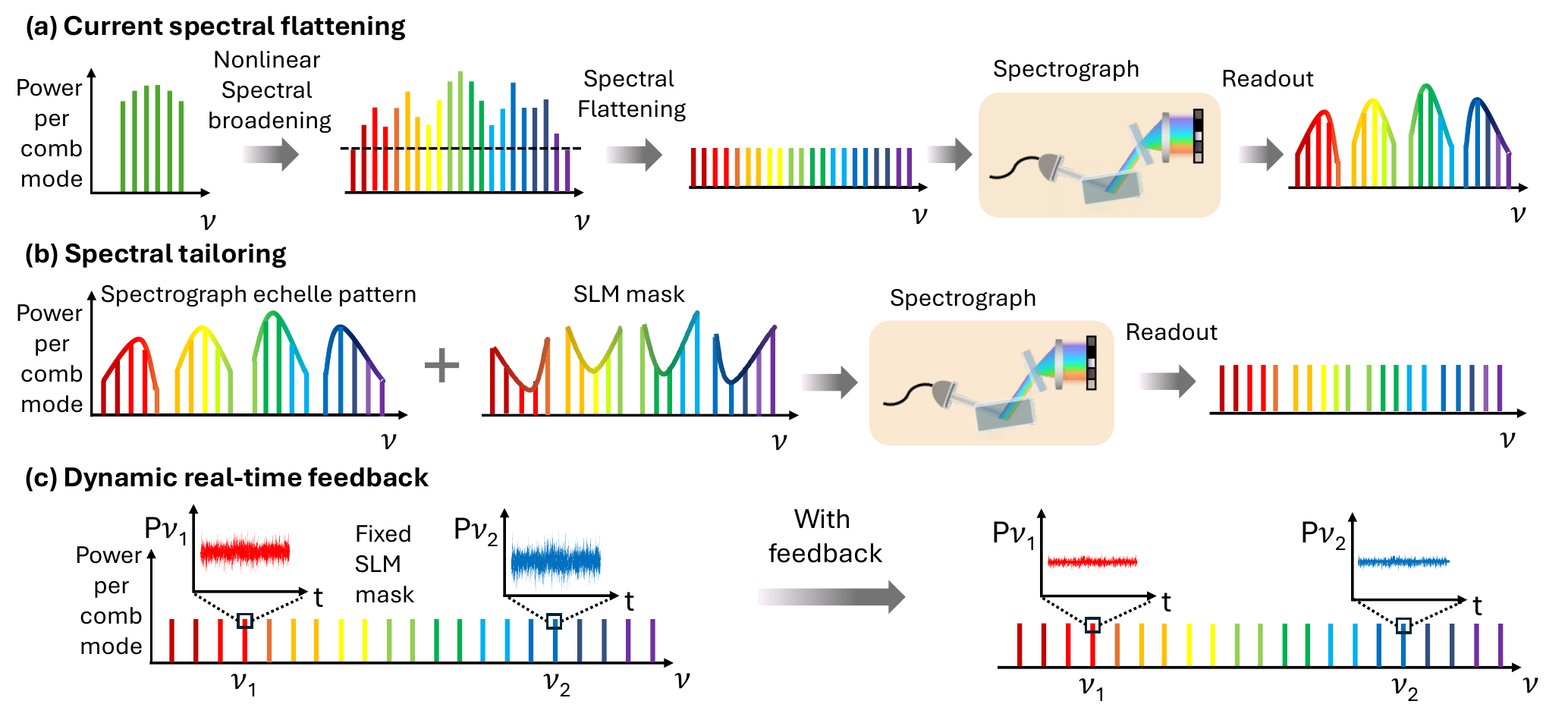}
\caption{Custom spectral tailoring and dynamic real-time feedback on an astrocomb. (a) Schematic illustrating the generation and spectral flattening of current state-of-the-art astrocombs. A narrowband high-repetition rate comb is generated from a filtered mode-locked laser, electro-optic modulator (EOM), or microresonator. This comb is spectrally broadened using highly-nonlinear fibers or nanophotonic waveguides, resulting in supercontinuum with inherent intensity variations, often exceeding factors of 20-30 dB across the wavelength band. A spectral flattener is implemented to equalize the flux of all comb lines across the entire spectral range of the detector. However, the detector's readout features a characteristic blazing pattern across different echelle orders and exhibits wavelength-dependent responsivity. (b) Schematic demonstrating custom spectral tailoring using a spatial light modulator (SLM). The SLM mask is tailored to compensate for the echelle pattern and the wavelength-dependent responsivity of the spectrograph, achieving a uniform flux across the detector on readout. (c) Illustration of the exaggerated effect of dynamic real-time flattening on the intensity variation of comb lines, highlighting the improvement in photon flux consistency.}
\label{fig1}
\end{figure}

Control of comb line intensities has been achieved using programmable attenuators such as liquid-crystal based spatial light modulators (SLM) \cite{probst2013spectral, debus2021spectral, serizawa2024laser}. Apart from astronomy, similar spectral control systems using SLMs are employed in temporal pulse shaping \cite{weiner2000femtosecond}, optical communications as a wavelength selective switch \cite{strasser2010wavelength} and in the pulse compression of frequency combs via higher-order dispersion compensation \cite{yamane2003optical}. SLMs are also extensively used in a variety of other applications, such as confocal microscopy for imaging biological systems \cite{ji2010adaptive}, arbitrary waveform generation \cite{cundiff2010optical}, programmable bandpass filtering \cite{ulman1993femtosecond}, 2D spectroscopy \cite{shim2009turn}, white light beam shaping \cite{spangenberg2014white} and steering for material processing \cite{yuan2020laser}, as well as in holography \cite{zhao2015accurate} and optical tweezers \cite{kim2016situ}, to cite a few. Compared to these methods, our spectral shaping technique offers both larger bandwidth and higher degree of intensity stability over extended periods. Recently, this technique of spectral flattening of laser frequency combs has also been integrated as an arrayed waveguide grating (AWG) on silicon nitride (SiN) and thin-film lithium niobate platform over narrower bandwidths around 1550 nm \cite{jovanovic2022flattening, wang2025electro}. 
 
Our immediate focus for this work relates to the 30 GHz electro-optic comb spanning 800 - 1300 nm that is used as the calibration source for the Habitat Zone Planet Finder spectrograph \cite{mahadevan2012habitable, mahadevan2014habitable, metcalf2019stellar}. The LFC at the HPF has been running reliably for the past six years, with an uptime exceeding 98$\%$, unparalleled by other astrocomb systems. However, the comb line intensities are controlled with a static spectral mask and can vary by as much as a factor of two with time. Additionally, the readout from the detector, common to all spectrographs, exhibits a characteristic echelle blazing pattern that depends on the optics and grating employed, as shown in Fig. \ref{fig1}(a). 

To address these shortcomings, here we implement custom spectral tailoring using an SLM that compensates for the echelle pattern, ensuring a constant flux level across the comb lines on readout from the detector, as illustrated in Fig. \ref{fig1}(b). Furthermore, we have demonstrated an improvement by approximately twofold in fractional intensity stability of our laboratory comb system through a dynamic real-time feedback loop to the SLM (Fig. \ref{fig1}(c)). Importantly, the impact of spectral flattening feedback has been evaluated through an out-of-loop measurement with an optical spectrum analyzer (OSA). Finally, we demonstrate these advances in spectral control across 800-1300 nm with an electro-optic comb pumped at 1.55 $\mu$m, instead of the ``in band" pumping at 1.064 $\mu$m that currently exists at the HPF.  This presents a significant advantage by reducing the required dynamic range for the spectral flattening to less than 12 dB, in contrast to over 20 dB with 1.064 $\mu$m pumping. In the future, we plan to employ these advantages to upgrade the performance of the calibration system at HPF.

\section{Spectral Flattening Setup}
\begin{figure}[ht!]
\centering\includegraphics[width=10cm]{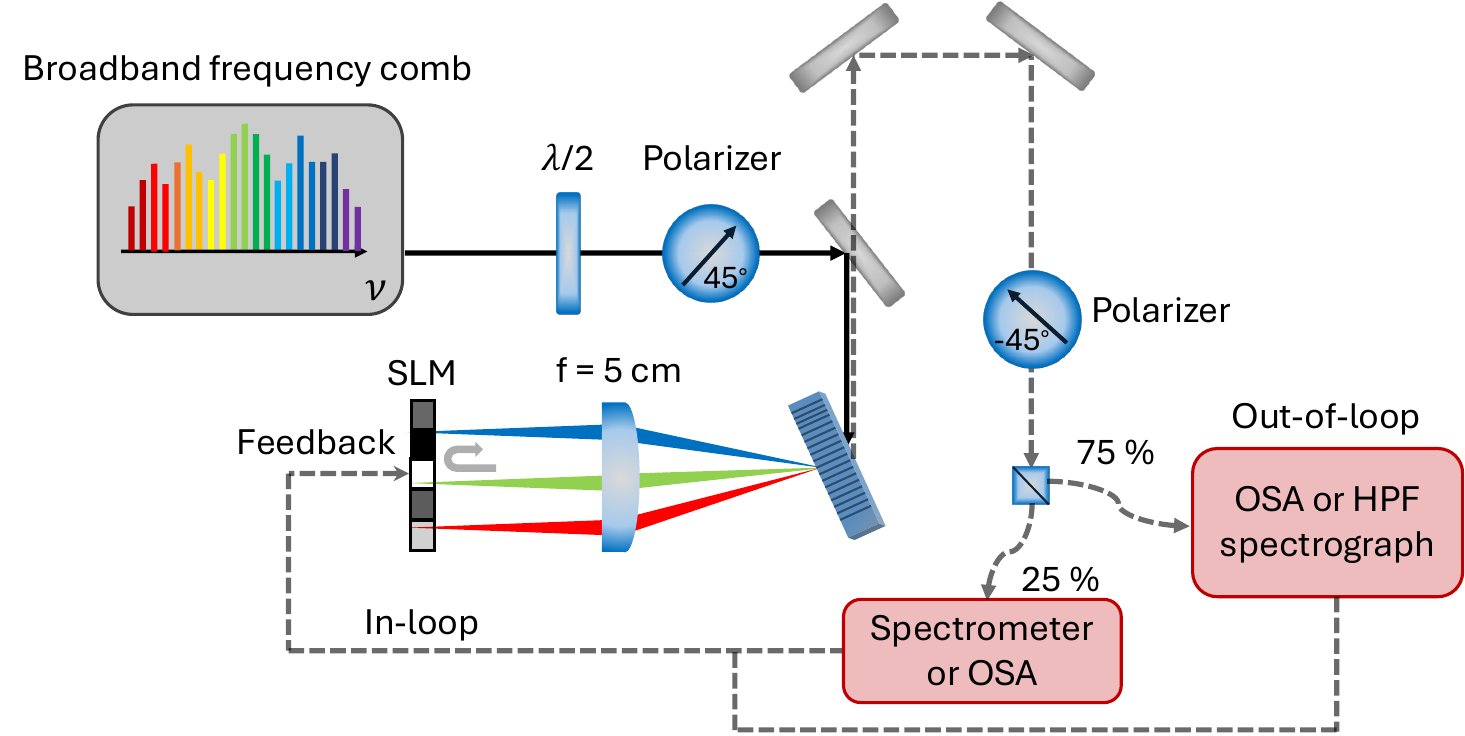}
\caption{Experimental setup for spectral flattening. $\lambda/2$: half-wave plate; SLM: spatial light modulator; OSA: optical spectrum analyzer; HPF: Habitable-zone planet finder spectrograph.}
\label{fig2}
\end{figure}
\begin{figure}[ht!]
\centering\includegraphics[width=1.0\linewidth]{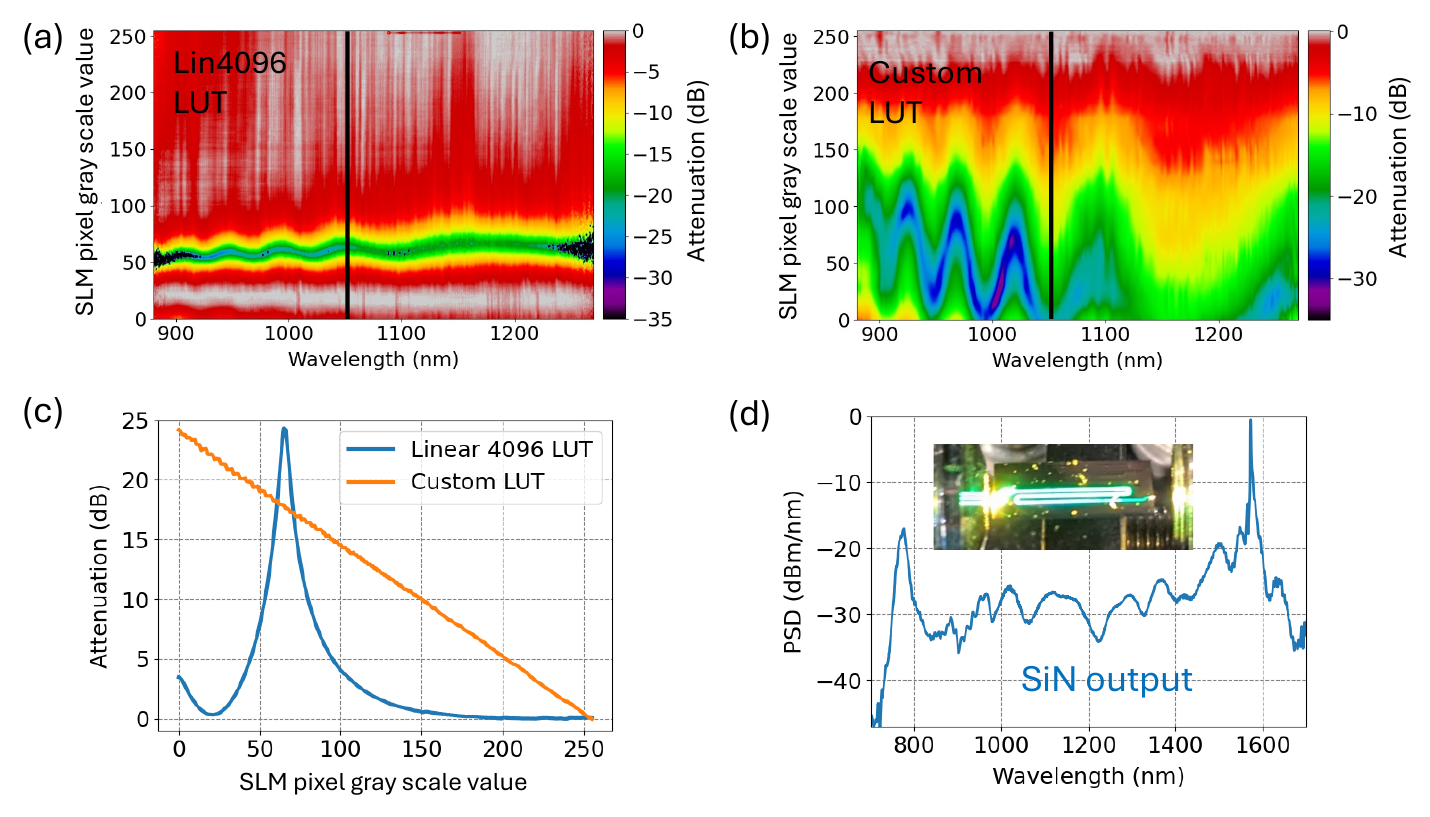}
\caption{Attenuation calibration of the spatial light modulator (SLM). The achieved attenuation range across the supercontinuum spanning 870 - 1280 nm on varying the grayscale value uniformly across all pixels of the SLM using, (a) linear look-up-table (LUT) varying from 0 to 4096 in increments of 16, and (b) Custom LUT designed for log-linear variation in attenuation from its minimum to maximum value across all wavelengths. (c) Attenuation at a central wavelength of 1064 nm using the above two LUTs. (d) Uncorrected supercontinuum output directly from the silicon nitride (SiN) waveguide pumped by a 10 GHz electro-optic comb at 1550 nm. Photo shows the top-view of the SiN chip generating the supercontinuum.}
\label{fig3}
\end{figure}

For this study, the LFC source is a 10 GHz resonant electro-optic comb pumped at 1550 nm. The comb is amplified to 5 W and temporally compressed to a duration of 50 fs. Further details on the comb system can be found in \cite{sekhar202320}. The compressed 10 GHz pulses are then launched into a hybrid SiN waveguide to generate a supercontinuum spanning 780 - 1700 nm as shown in Fig. \ref{fig3}(d). The SiN waveguide is 23.5 mm long with a thickness of 800 nm. It features a hybrid design consisting of an initial 15 mm long segment with a width of 2.6 $\mu$m followed by a section that is 8.5 mm long and 4.8 $\mu$m wide. This hybrid dispersion profile along the waveguide's propagation length facilitates the generation of a broadband and relatively flat spectrum \cite{carlson2019generating, patent}. Additionally, pumping at 1550 nm significantly reduces the dynamic range required for spectral flattening to less than 12 dB across the entire HPF spectral bandwidth of 800 - 1300 nm, as the pump peak at 1550 nm lies outside the desired wavelength range. This is in contrast to previous systems that had their pump within the desired wavelength range and required a larger dynamic range for spectral flattening \cite{probst2013spectral, debus2021spectral, metcalf2019stellar}. 

Fig. \ref{fig2} is a schematic of our experimental setup. This approach to Fourier-domain optical pulse shaping is similar to previous works, such as \cite{probst2013spectral, debus2021spectral, metcalf2019stellar, zhen2024improving}. The broadband frequency comb is collimated and polarized at 45$^{\circ}$ with respect to the liquid crystal axis of the SLM using a half-wave plate and a linear polarizer. The beam is then dispersed by a transmissive volume holographic grating and the individual spectral components are focused onto the SLM using a cylindrical lens of 5 cm focal length ($f$). The reflective spatial light modulator (SLM) used in this study is a Meadowlark model: E19x12-600-1300-OH, with 1920 $\times$ 1200 pixels covering an active area of 15.36 $\times$ 9.60 mm. The reflected light from the SLM is re-collimated and recombined by the cylindrical lens and grating respectively. Both the grating and SLM are positioned at a distance $f$ from the cylindrical lens, forming a 4$f$ configuration \cite{weiner2000femtosecond}. The broadband frequency comb is spread across the longer edge of the SLM achieving a spectral resolution of 0.4 - 0.6 nm across the wavelength range of 800 - 1300 nm. For additional details on the spectral flattening setup, refer to Figure S1 in the Supplement.

The voltage applied to each pixel of the SLM induces a rotation of the liquid crystal. This adds a phase shift along one polarization axis, thereby allowing the SLM to function as a variable waveplate. A cross-polarizer at the output converts the polarization rotation induced by the SLM into a wavelength-dependent attenuation. The output light is then split using a fiber-integrated 75:25 multimode splitter with a core diameter of 50 $\mu$m. The 1m long fiber splitter transmits $> 99\%$ of the input light across 650 - 2000 nm. The 25$\%$ port is sent to a grating-based spectrometer with an InGaAs linear array to monitor the output spectrum and provide feedback to the SLM for real-time intensity adjustments. The remaining 75$\%$ port is used to measure the flattened LFC in an "out-of-loop" optical spectrum analyzer (OSA). When used at the telescope site, the OSA would be replaced by the astronomical spectrograph.    
\section{Calibration of SLM}
The spectral flattening technique using an SLM involves two major calibration steps: (i) Attenuation calibration, which maps the power transmitted at a particular wavelength as a function of the voltage applied to the SLM pixels, and (ii) Wavelength calibration, which maps the pixel column number to a specific wavelength. The section below describes both calibration procedures in detail.

\subsection{Mapping attenuation to grayscale value on SLM}
In order to setup the spectral flattening system and perform the attenuation calibration, we start with a continuous wave (CW) laser at the center wavelength of 1064 nm. The first polarizer is aligned at 45$^{\circ}$ with respect to the liquid crystals, while the second polarizer is cross-polarized to minimize the power transmitted through the combined polarizer system. The spectrometer port of the 25/75 beamsplitter shown in Fig. \ref{fig2} is replaced by a powermeter to measure the transmitted power. %It is essential to ensure that the powermeter is zeroed for background light before starting the measurements. 

The SLM has a 12-bit controller. There are 4096 possible voltage levels that can be applied to the liquid crystals to achieve varying degrees of phase shift. However, the computer-display interface of the device is limited to 8-bits per pixel, so a look up table (LUT) is used to map the 256 grayscale values to discrete voltage levels. Only 256 of the 4096 values are accessible through any one LUT. The main purpose of the attenuation calibration is to choose a LUT that gives the greatest dynamic range and resolution.
%The SLM has an inherent look-up-table (LUT) that maps the input grayscale values from 0 to 255 (for the 8-bit system) to a corresponding set of output values, resulting in a linear phase shift from 0 to $2\pi$.
In this system, the LUT referred to as `Lin4096' ranges from 0 to 4096 in steps of 16. Recalibration of the LUT is necessary to achieve the desired phase shift for most applications.
% Trans vs phase shift eqn?, flat exposure

To optimize the polarization of input light to the SLM for maximum dynamic range of attenuation, we follow the procedure outlined below. First, we set the grayscale value uniformly across the SLM pixels to the value that results in the lowest transmitted power. Next, we fine-tune both the output and input polarizers to further minimize the transmitted power measured. The half-wave plate is then adjusted to align the input light polarization with the axis of the first polarizer. Following these adjustments, we gradually change the grayscale value across the SLM pixels from 0 to 255, measuring the transmitted power at each step and recording the dynamic range achieved. These steps of tuning the SLM grayscale scale, adjusting the polarizers, and fine-tuning the half-wave plate are repeated until the maximum dynamic range is obtained.

With the optical components of the system fixed, the CW laser and powermeter are replaced by the broadband LFC and OSA respectively. The grayscale value is varied from 0 to 255 using the Lin4096 LUT and the resulting power variation of the supercontinuum across 870 - 1280 nm is recorded on the OSA. The maximum intensity at each wavelength is noted and the relative attenuation across the entire wavelength range is presented in Fig. \ref{fig3}(a). The results indicate that the net phase shift wraps through $2\pi$, demonstrating that the Lin4096 LUT does not provide a linear attenuation with respect to the pixel value. This LUT covers a phase shift larger than required and does not allow small step sizes as needed to make fine adjustments in intensity. 

Consequently, we generate a custom LUT such that the grayscale values 0-255 correspond to the minimum and maximum transmitted power at the center wavelength of 1064 nm. The attenuation map generated using this custom LUT is displayed as a false color image in Fig. \ref{fig3}(b). The attenuation achieved at 1064 nm using both the LUTs is also plotted in Fig. \ref{fig3}(c). The system achieves a maximum attenuation of 25 dB at 1064 nm and even reaches 30 dB at 1 µm, with step sizes as low as 0.09 - 0.1 dB. Since liquid crystals are not achromatic, the induced phase shift is wavelength-dependent as shown in Fig. \ref{fig3}(b). As a result, we record the grayscale values corresponding to the minimum and maximum transmitted power at each wavelength and use that information to bound the range of the spectral flattening algorithm at other wavelengths.
\begin{figure}[ht!]
\centering\includegraphics[width=\linewidth]{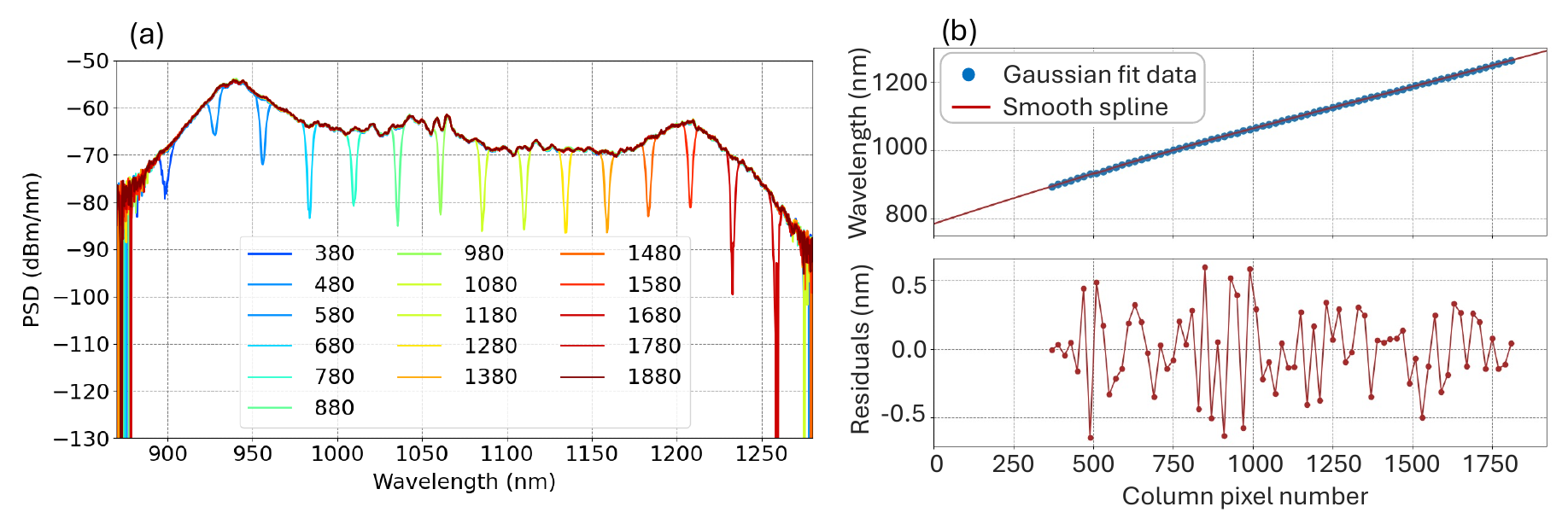}
\caption{Wavelength-to-pixel calibration of SLM. (a) The SLM is configured for maximum attenuation column-wise in steps of 20. Legends indicate the column numbers. The resulting transmission spectra from the spectral flattener recorded by the optical spectrum analyzer (OSA) at each step are overlaid. (b) The blue dots in the top plot show the center wavelengths obtained by fitting a Gaussian to the corresponding trasmission spectra from (a). The red line indicates a smooth spline fit to the center wavelengths, while the bottom plot displays the residuals between the smooth spline and the Gaussian fit centers.}
\label{fig4}
\end{figure}
\subsection{Mapping wavelength to SLM pixel column  number}
Wavelength calibration is performed to identify the center wavelength corresponding to each column pixel number of the SLM, which is perpendicular to the axis along which input light is dispersed. Initially, all the pixels are uniformly set to 255, corresponding to nearly maximum transmitted power across all wavelengths and the measured spectrum is recorded. Then a set of 20 pixel columns are sequentially set to zero to achieve maximum attenuation. In this way, the entire SLM is mapped in steps of 20, with the transmission spectra recorded at each step as shown in Fig. \ref{fig4}(a). 

The difference between the transmitted spectra at each step and the spectrum with no attenuation is calculated and fitted with a Gaussian function. The center wavelength is then attributed to the mean of the column pixels set to maximum attenuation, represented by the blue dots in Fig. \ref{fig4}(b). The measured wavelength-to-pixel data is fitted with an extrapolated smooth spline (red trace in Fig. \ref{fig4}(b)), which is used as the wavelength calibration curve for the spectral flattening algorithm. The difference between the measured and fitted data is within 0.5 nm, which is less than the full-width at half-maximum (FWHM) of the optical resolution of the spectral flattening setup as shown in the supplement Fig. S3. 

This precise calibration is necessary to prevent intensity adjustments at a neighboring wavelength instead of the desired one, which can result in undesirable modulations on the LFC envelope during prolonged operation of the spectral flattening algorithm. This wavelength calibration procedure has been performed separately using both the OSA and spectrometer. 

\section{Demonstration of real-time spectral shape tailoring}
% grating efficiency bandwidth issue - doesn't cover fully 800 - 1300 nm. 
The goal of the spectral flattening setup is to maintain a constant photon flux for all comb modes across the spectrograph bandwidth, which is 800 - 1300 nm for the HPF. The spectral flattening output is measured on an OSA or spectrometer in terms of spectral density (in mW per nm or counts per nm respectively). The number of comb lines per nanometer is inversely proportional to the square of the wavelength, and the photon energy is proportional to the inverse of wavelength. 
Thus, a constant power spectral density corresponds to the photon flux around a comb mode being directly proportional to the third power of the wavelength. This is the photon flux (number of photons/sec) per comb mode available for the astronomical spectrograph.

To compensate this, the reference set in the SLM has a negative slope with respect to wavelength as illustrated by the black dashed line in Fig. \ref{fig5}(a). The original grating employed for the results presented in Fig. \ref{fig6} and Fig. \ref{fig7} has 600 lines per mm and diffraction efficiency over a narrow bandwidth. To address this, it was replaced with a grating, with 500 lines per mm, that operates across the full 800–1300 nm bandwidth, as shown in Fig. \ref{fig5}. A comparison of the insertion loss (IL) for the spectral flattening setup, using both gratings, is presented in the bottom plot of Fig. \ref{fig5}(b). The measured insertion loss is without including any additional attenuation from the SLM. 

\subsection{Spectral flattening feedback loop}
\begin{figure}[ht!]
\centering\includegraphics[width=\linewidth]{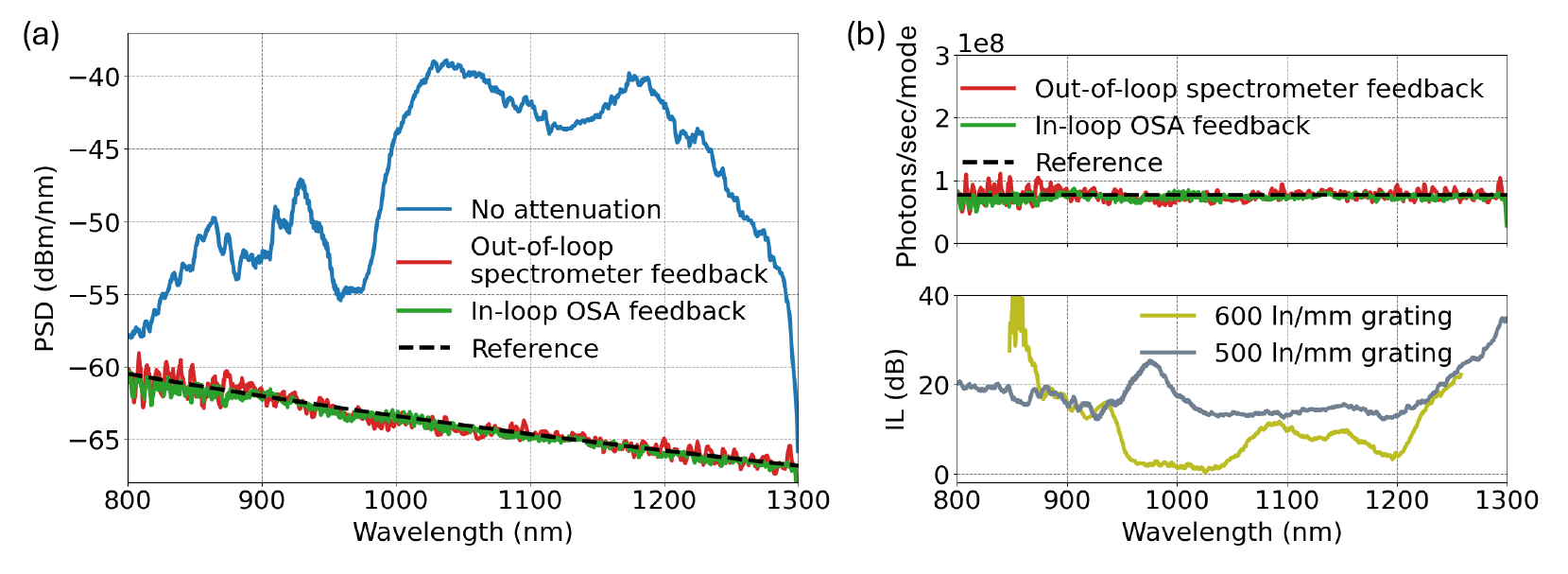}
\caption{Measured spectrum of a 10 GHz laser frequency comb (LFC) spectrum on the out-of-loop OSA after spectral flattening. (a) Power spectral density (PSD in dBm/nm) of the output LFC from the spectral flattener, comparing results with and without applying attenuation. (b) Constant photon flux per comb mode (photons/sec/mode in linear scale) of the spectral flattener outputs from (a). The out-of-loop measurement (red curve) reflects the LFC output recorded on the OSA, while feedback to the spatial light modulator (SLM) is obtained from the spectrometer. In contrast, for the in-loop measurement (green curve) both the output and feedback to the spectral flattener are provided by the same instrument, the OSA. The bottom plot compares the insertion loss (IL) of the spectral flattening setup using the two gratings.}
\label{fig5}
\end{figure}

Using the wavelength-dependent attenuation-to-pixel value calibration and wavelength-to-pixel number map, a spectral flattening feedback algorithm was developed in Python. The process begins by recording the spectrum on the in-loop spectrometer without any spectral shaping, as shown by the blue trace in Fig. \ref{fig5}(a). The algorithm calculates the difference between the measured spectrum and the target (reference) spectrum across the supercontinuum spanning 800 - 1300 nm. Employing the two calibration curves of Fig. \ref{fig3}(b) and Fig. \ref{fig4}(b),  the algorithm applies the pixel value corresponding to the required attenuation at each specific wavelength to the particular pixel column in the SLM. It is crucial to restrict the range of grayscale pixel values at each wavelength to its minimum and maximum attenuation range to ensure the feedback maintains the correct sign. 

A single iteration achieves an intensity level within 3 dB of the target.  This procedure is repeated in a feedback loop, with a new spectrum recorded after each iteration. The updated wavelength-dependent attenuation is recalculated, and the corrected grayscale pixel values are applied to the corresponding column pixels. The spectrum measured on the OSA reaches a steady state in less than 10 iterations.

\subsection{In-loop vs out-of-loop measurements}
The resulting flattened spectrum represented by the green and red traces in Fig. \ref{fig5}(a) are obtained when feedback is applied from the OSA and the spectrometer respectively. We refer to these two cases as in-loop or out-of-loop measurements, depending on whether the feedback and flattened spectrum are measured using the same instrument or two different instruments. This distinction is important to make and test. In actual use with an astronomical spectrograph (such as the HPF), the calibration spectrum from the HPF will be out-of-loop and acquired on a much longer timescale than the in-loop spectrometer-based feedback.  

In the case of an out-of-loop measurement, a precise intensity cross-correlation measurement needs to be performed between the two instruments to establish a new target spectrum in the SLM. A single iteration takes approximately 1.6 seconds when feedback is applied from the OSA, while it is an order of magnitude faster (around 0.3 seconds) using the spectrometer for feedback. However the degree of flatness achieved is within 1.6 dB for the in-loop measurement and 2.6 dB for the out-of-loop measurement across the wavelength range of 860 - 1300 nm. Figure \ref{fig5}(b) presents these measurements on a linear scale, displaying the flattened photon flux per comb mode. This clearly illustrates that the degree of flatness for the out-of-loop measurement is not as small as that for the in-loop measurement, with more high-frequency intensity fluctuations observed. These fluctuations are related to the accuracy of the intensity cross-calibration between the OSA and the spectrometer, which includes wavelength-dependent differences in responsivity of the two. The accuracy is further limited by the lower sensitivity of the spectrometer compared to the OSA and the accuracy of wavelength calibration. For additional details, refer to Figure S2 in the Supplement.
\begin{figure}[ht!]
\centering\includegraphics[width=\linewidth]{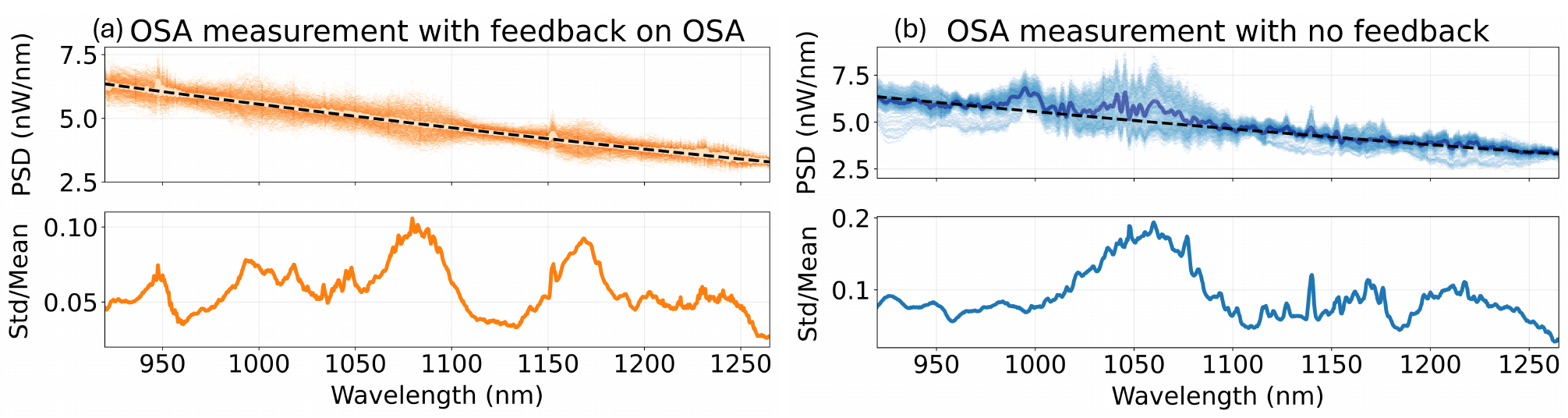}
\caption{Comparison of the flattened 10 GHz laser frequency comb (LFC) spectrum on an optical spectrum analyzer (OSA) after the spectral flattener: (a) with, and (b) without real-time feedback. The top plots illustrate the variation in power spectral density of the comb over a one-hour measurement period, while the bottom plots show the corresponding standard deviation by mean values. Notably, the implementation of real-time feedback reduces the fractional intensity fluctuations by almost a factor of two.}
\label{fig6}
\end{figure}
\subsection{Impact of real-time feedback on spectral stability}
The 10 GHz LFC in this laboratory setup is not frequency stabilized to the GPS and the power levels at each step of broadening are not monitored in contrast to the system installed with the HPF at the Hobby-Eberly telescope \cite{metcalf2019stellar}. The only feedback control in the experimental setup, aside from the SLM, is to the nanomax stage, which is actively piezo-controlled to maintain the same coupling efficiency to the SiN waveguide with the high average powers during measurements (typically 3 W). We observe that  fractional intensity fluctuations at certain wavelengths of the supercontinuum can reach as high as 20$\%$. Fig. \ref{fig6} compares the stability of power spectral density of the supercontinuum across 910 - 1260 nm with (a) and without (b) real-time feedback from the OSA. The traces in the top plots are recorded every 10 seconds over a duration of one hour, while a single feedback iteration takes approximately 1.6 seconds. 

The results indicate that real-time feedback reduces fractional intensity fluctuations by a factor of two. The mean value of fractional intensity fluctuations is 5$\%$ in the presence of real-time feedback, which may be limited by the technical noise of the measurement system, not stabilized within the feedback iteration time. This may not pose a problem for actual observations at the HPF or another telescope site, where stellar and calibration spectra are typically acquired with an integration time of 5-10  minutes. The Allan deviation of intensity fluctuations at an averaging time of 10 minutes is nearly an order of magnitude lower than at 10 seconds in the presence of real-time feedback as shown in the Supplement. The stability of feedback loop and the degree of spectral flatness were also monitored for over 6 hours, showing no signs of degradation.  
\begin{figure}[ht!]
\centering\includegraphics[width=\linewidth]{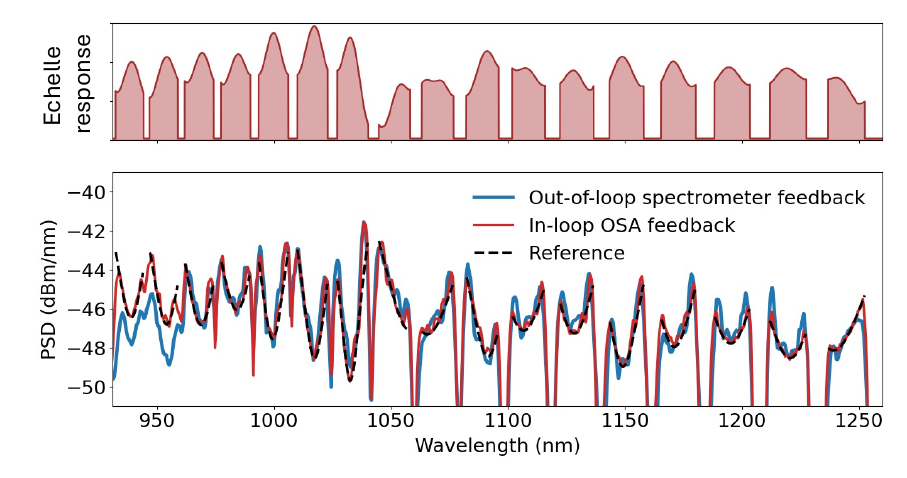}
\caption{Compensation of spectrograph response. The top plot represents a fitted model of the echelle response data from the HPF spectrograph. The bottom plot shows the measured output of the laser frequency comb (LFC) from the spectral flattener on an optical spectrum analyzer (OSA) after applying a spatial light modulator (SLM) mask to compensate for the blazing pattern of different echelle orders in the HPF spectrograph. The blue and red traces on the bottom plot represent the feedback to the SLM obtained from a spectrometer and OSA respectively.}
\label{fig7}
\end{figure}
\subsection{Compensation of the echelle blazing pattern of an astronomical spectrograph response}
Every astronomical spectrograph exhibits a characteristic blazing pattern that depends on the grating and optics used. Each echelle order has a specific bandwidth, with gaps in the spectral readout between different orders. The top plot in Fig. \ref{fig7}(a) shows a fit to the typical echelle response from the HPF spectrograph. As a result, even a flattened LFC spectrum shown in Fig. \ref{fig5}(b) does not translate to a constant flux per comb mode in the spectrograph readout. In this work, we utilize the spectral flattener described above in Fig. \ref{fig1} to tailor the LFC spectrum in real-time, compensating for this echelle response to achieve a constant LFC flux per comb mode in the spectrograph readout. 

The in-loop (red trace) and out-of-loop (blue trace) measurements of this inverse echelle response (dashed black trace) are plotted in Fig. \ref{fig7}(b). The results demonstrate a good degree of agreement with the target curve. The disagreement between the out-of-loop measurement and the target may be attributed to the intensity cross-calibration and the lower sensitivity of the spectrometer as compared to OSA (see Fig. S2 in supplement). This approach may also significantly reduce the impact of scattered comb light on the stellar spectrum by attenuating the comb lines between the echelle orders.

% \begin{figure}[ht!]
% \centering\includegraphics[width=\linewidth]{Fig_sens_resn.pdf}
% \caption{Comparison of sensitivity and resolution of the OSA and spectrometer. (a) Sensitivity of OSA and spectrometer with no input light. The blue trace represents the full width at half-maximum (FWHM) of the instrumental line shape (ILS) on activating a single pixel column of SLM using (b) OSA (at a set resolution of 0.02 nm) and (c) spectrometer. The red dashed line shows the mean of the FWHM of the ILS.}
% \end{figure}

\section{Conclusion}

In summary, we have demonstrated a dynamic spectral flattening setup for astrocombs, using a spatial light modulator, as a critical aspect of radial velocity spectroscopy aimed at achieveing cm/s RV precision. In contrast to the current LFC setup at HPF that spans 800 - 1300 nm, this work provides two key advancements. First, custom spectral tailoring is used to compensate for the characteristic echelle pattern of the spectrograph. This tailoring helps maintain a constant flux readout of the LFC from the detector, while also reducing the contamination of scattered comb light on the stellar spectrum by attenuating the comb light between the echelle orders. Second, real-time feedback to the SLM mitigates spectral intensity variations over time, improving the spectral stability by nearly a factor of two. 

An additional important aspect of our work is the analysis of the intensity cross-calibration between in-loop and out-of-loop spectral flatness measurements. This provides the basis for future implementation and improvement on actual astronomical spectrographs. Our approach further reduces the dynamic range required for spectral flattening by pumping at a wavelength outside the desired calibration range, effectively nullifying the impact of high-frequency pump fluctuations on spectral stability. Finally, in addition to its applications in astronomy, similar spectral shaping setups with broad spectral coverage and high resolution have potential uses in telecommunications, 2D spectroscopy, material processing, and optical tweezers.  

\begin{backmatter}
\bmsection{Funding}
National Science Foundation (AST 2009982)

%\bmsection{Acknowledgment}
%We thank Michael Geiselmann of Ligentec for providing the SiN waveguides. 

\bmsection{Disclosures}
The authors declare no conflicts of interest. 

\bmsection{Data Availability Statement}
Data underlying the results presented in this paper may be obtained from the authors upon reasonable request.

\bmsection{Supplemental document}
See Supplement 1 for supporting content.

\end{backmatter}

%%%%%%%%%%%%%%%%%%%%%%% References %%%%%%%%%%%%%%%%%%%%%%%%%

%%%%%%%%%% If using BibTeX:
\bibliography{sample}

\end{document}

\maketitle

\section{Designing the spectral flattener setup}

The employed spatial light modulator (SLM) has 1920 $\times$ 1200 pixels, each of size 8 $\mu$m $\times$ 8 $\mu$m, covering an active area of 15.36 $\times$ 9.60 mm. The spectral flattening setup is illustrated in Fig. 2 of the main manuscript. The entire setup, excluding the spectrometer and OSA, fits on a 6" $\times$ 24" breadboard. 

This section provides additional details on the design of the spectral flattener for achieving the desired spectral shaping \cite{monmayrant2010newcomer}. The supercontinuum input to the spectral flattener is collimated using an off-axis parabolic (OAP) mirror, resulting in a beam radius $w_{i}$. The off-axis parabolic mirror reduces chromatic aberration on collimating a broadband supercontinuum. The beam is then polarized and dispersed by a transmissive volume holographic grating. The grating is chosen such that it offers maximum efficiency across the desired wavelength range of 800 - 1300 nm. With the grating being fixed, the angular diffraction range is determined by the grating equation,
\begin{equation}
d\left(\sin \theta_{i}+\sin \theta_{d}\right)=m \lambda,
\label{grating_eqn}
\end{equation}
where $d$ is the grating constant, $\theta_{i}$ and $\theta_{d}$ are the grating incident and diffraction angles respectively. For this application, we consider only the first order diffraction (m = 1). The incident angle is maintained close to the Bragg angle at the center wavelength (1064 nm) of the supercontinuum.

To achieve the maximum spectral resolution in the setup, the spectral range ($\Delta \lambda$) of 500 nm must be spread across the longer edge of SLM. The required angular diffracted range is given by,
\begin{equation}
 \Delta \theta_{d}=\Delta x / f,
 \label{x_range}
\end{equation}
where $\Delta x$ is the SLM aperture along the longer edge (15.36 mm for the SLM used in this study) and $f$ is the focal length of the cylindrical lens. In order to achieve this, we can determine the required focal length of the cylindrical lens by differentiating equation \ref{grating_eqn} and substituting the result into the equation \ref{x_range} yielding,
\begin{equation}
    f=d \cos \theta_{d}   \Delta x / \Delta \lambda.
\end{equation}

\begin{figure}[htbp]
\centering
\centering\includegraphics[width=\linewidth]{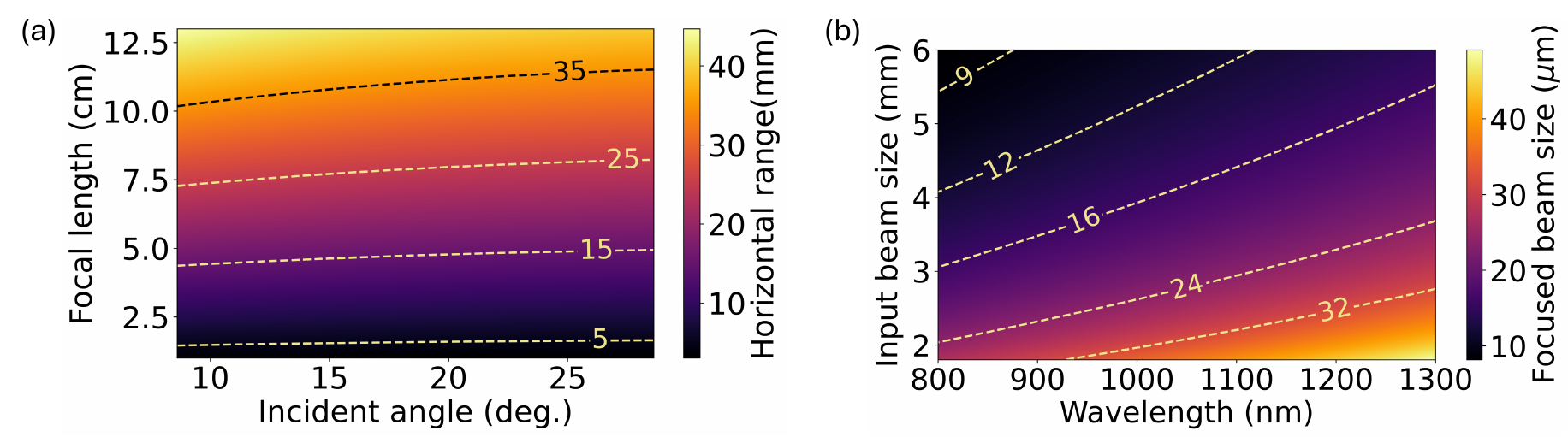}
\caption{Design of the spectral flattener: (a) Spatial range of the diffracted beam on the SLM along the dispersion direction as a function of the incident grating angle and focal length of the cylindrical lens. (b) Focused beam size on the SLM across 800–1300 nm wavelength range, on varying the input beam size while using a fixed cylindrical lens with a focal length of 5 cm.}
\label{design}
\end{figure}

Figure \ref{design}(a) shows the achievable spatial range of the diffracted beam on the SLM as a function of the grating incident angle and focal length of the cylindrical lens. A focal length of approximately 5 cm provides a spatial range close to 15 mm, matching the longer edge of the SLM and thus, allowing majority of the SLM pixels to be illuminated. The focused beam size of each spectral component ($\lambda$) on the SLM is calculated using Gaussian optics as given by,  
\begin{equation}
    w_{o}=\frac{\lambda f}{\pi w_{d}},
\end{equation}
where $w_{o}$ and $w_{d}$ correspond to the beam radii of the focused and diffracted beams respectively. The beam radius of diffracted beam can be calculated using geometric optics, given by $w_{d}=\frac{w_{i} \cos \theta_{d}}{\cos \theta_{i}}$. Figure \ref{design}(b) shows the calculated focused beam size across the desired range of 800 - 1300 nm for a fixed cylindrical lens with a focal length of 5 cm and varying input beam diameters ($2 w_{i}$) between 2 and 6 mm. From this analysis, we selected an intermediate collimation size of 4 mm to achieve a focused beam spot size of 12 - 22 microns across the spectral range. This corresponds to approximately 2 - 3 pixels on the SLM. Thus the calculated resolution of the SLM is $\sim$ 0.5 nm per pixel. %, which is close to the resolution of the OSA. 

\section{Sensitivity and Resolution}

\begin{figure}[htbp]
\centering
\centering\includegraphics[width=0.7\linewidth]{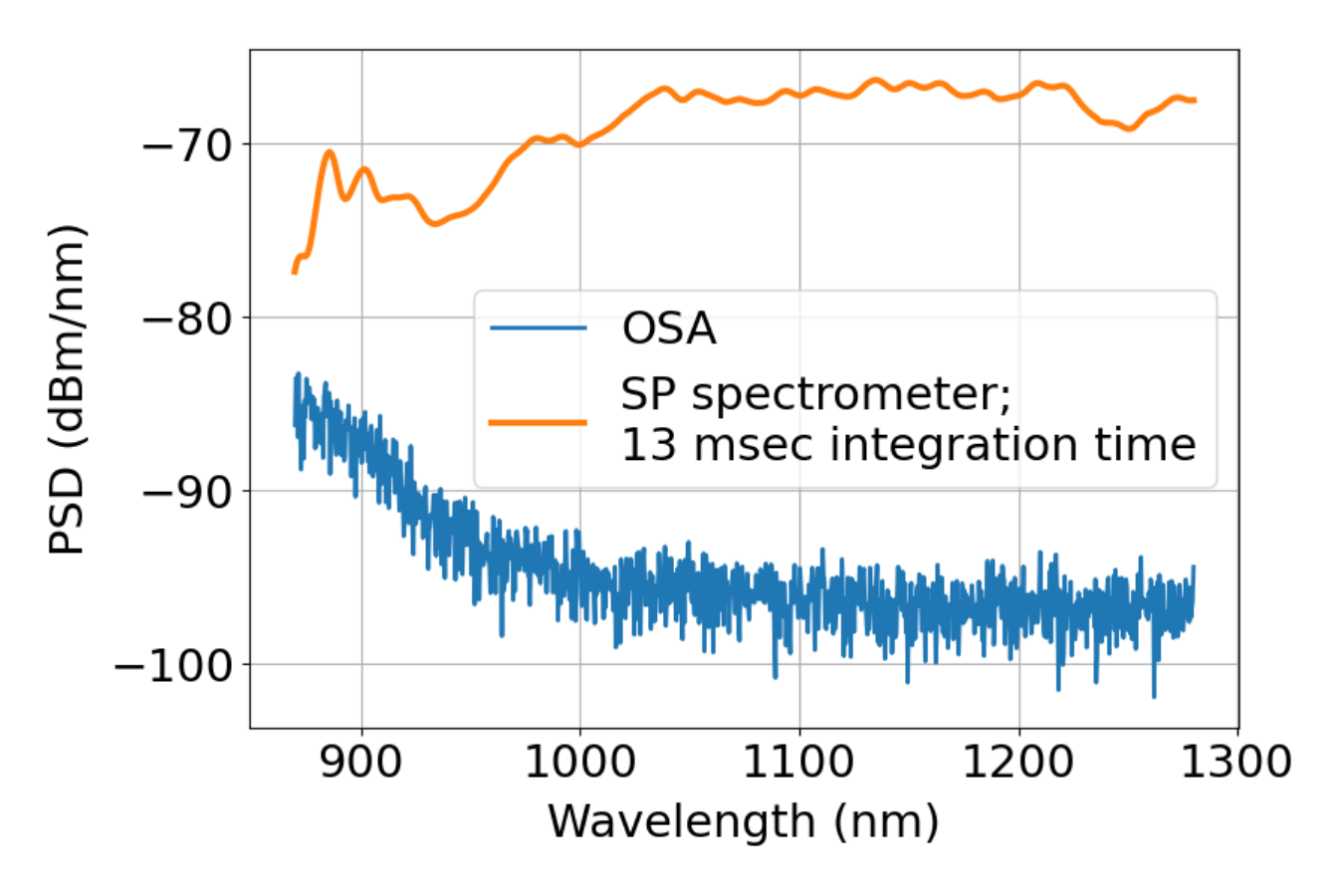}
\caption{Comparison of sensitivity of the OSA and spectrometer. The measurement is done with no input light to the instrument. OSA settings: sensitivity mode of HIGH1 and 2 nm resolution. Spectrometer measurement at an integration time of 13 msec.}
\label{sens}
\end{figure}

In this study, we used two instruments - an OSA (Yokogawa model: AQ6370D) and spectrometer (Spectral Products model: SM304-512) - to provide feedback and analyze the output from the spectral flattener through both in-loop and out-of-loop measurements. In the out-of-loop measurement, feedback to the SLM is obtained from the spectrometer while spectral flattener output is recorded on the OSA. Here, high-frequency intensity fluctuations are observed to be greater than those in the in-loop measurement (see Fig. 5 in the main manuscript). This discrepancy can be attributed to the differences in sensitivity between the OSA and the spectrometer, which limits the accuracy of intensity cross-calibration. The OSA operates in HIGH1 sensitivity mode with a resolution setting of 2 nm. This resolution setting is chosen to enable fast feedback to the SLM, in the order of $\sim$ 1 second. The spectrometer data is recorded with an integration time of 13 msec, the same duration used in the actual spectral measurements to prevent detector saturation at any wavelength. Fig. \ref{sens} shows that the sensitivity of OSA is approximately two orders of magnitude better than that of the spectrometer, along with a higher dynamic range before saturation. This noise floor measurement is done with no input light and employs the intensity cross-calibration data obtained between the two instruments, taking into account the difference in their wavelength-dependent responsivities.
\begin{figure}[htbp]
\centering
\centering\includegraphics[width=\linewidth]{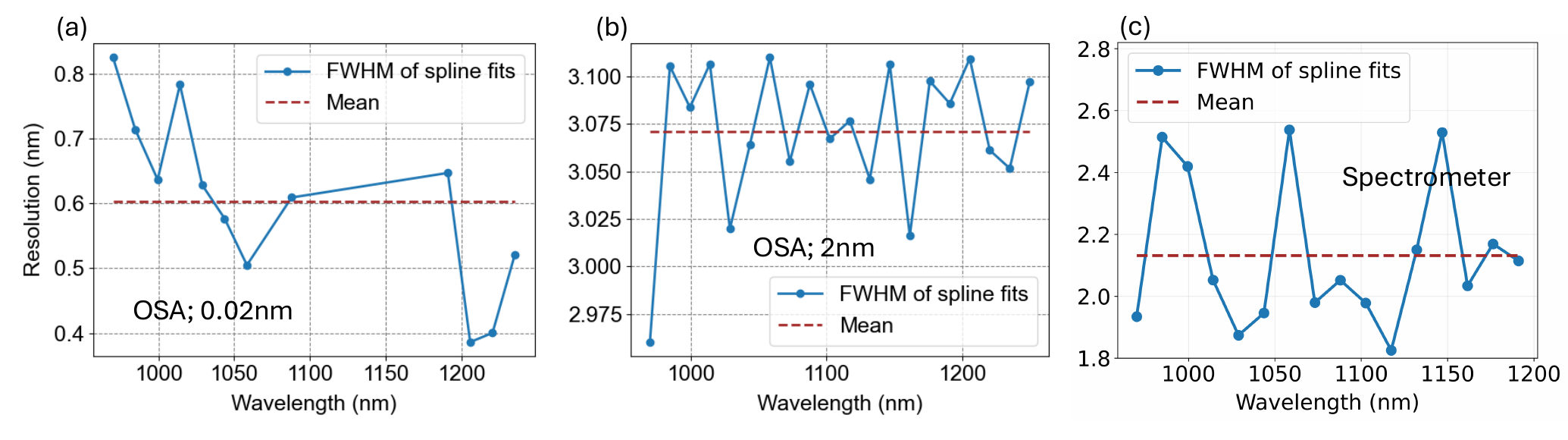}
\caption{Comparison of resolution of the OSA and spectrometer. The blue trace represents the full width at half-maximum (FWHM) of the measured instrumental line shape (ILS) on activating a single pixel column of SLM using (a) OSA at a set resolution of 0.02 nm, (a) OSA at a set resolution of 2 nm which is used during most of the measurements and (c) spectrometer. The red dashed line shows the mean of the FWHM of the ILS.}
\label{res}
\end{figure}

Fig. \ref{res} show the full-width at half-maximum (FWHM) of the instrumental line shape (ILS) measured using the OSA and the spectrometer. This measurement is performed by activating a single pixel column of the SLM, measuring its readout on the respective instrument and fitting a smoothed spline to the transmission spectrum. Fig. \ref{res}(a) and (b) represent the FWHM measured on the OSA with resolution settings of 0.02 nm and 2 nm respectively. The ILS has a mean FWHM of $\sim$ 0.6 nm (from Fig. \ref{res}(a)), which is close to the calculated resolution of the SLM presented in the previous section. The mean resolution of the ILS across 970 - 1250 nm in the measurements is found to be 3.07 nm for the OSA and 2.13 nm for the spectrometer as shown in Fig. \ref{res}(b) and (c), limited by the resolution of the respective instruments.  

\section{LFC intensity  stability}
\begin{figure}[htbp]
\centering
\centering\includegraphics[width=\linewidth]{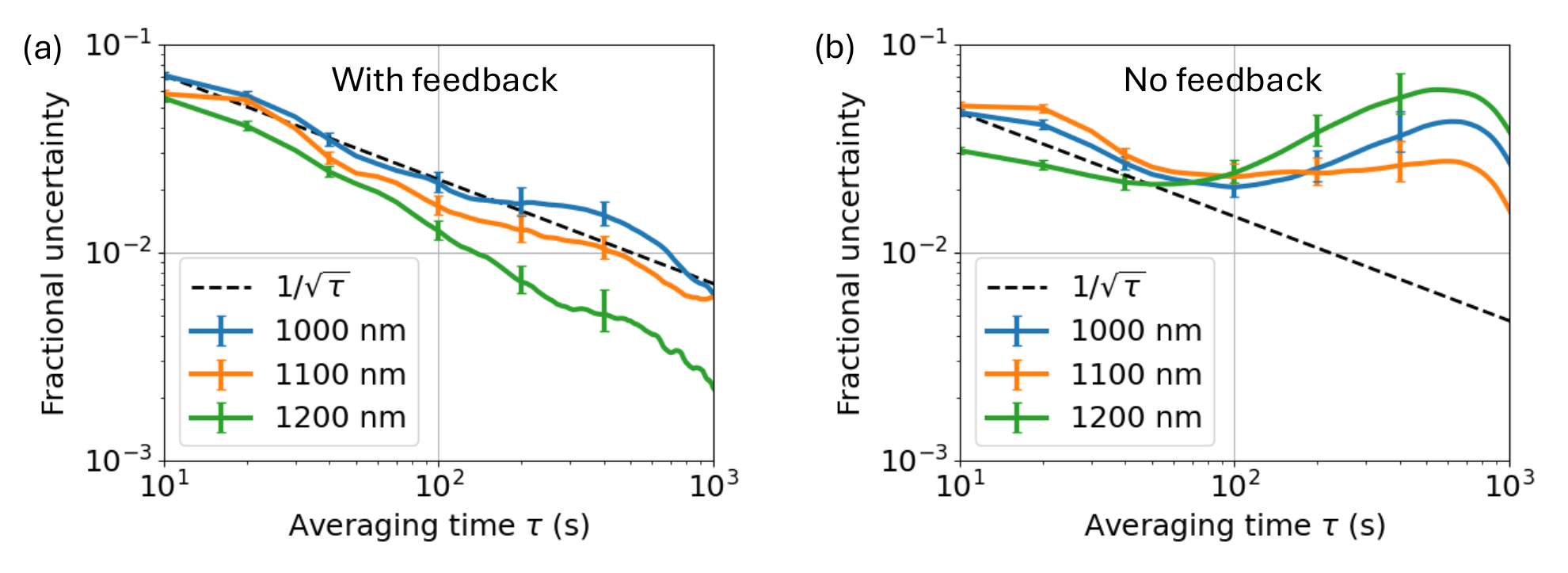}
\caption{Intensity stability of the flattened frequency comb. Allan deviation of intensity fluctuations at three different wavelengths of 1000 nm, 1100 nm and 1200 nm: (a) with, and (b) without real-time feedback. The black dashed line represents the $1/\sqrt{\tau}$ scaling of uncertainty.}
\label{adev}
\end{figure}

Figure \ref{adev} shows the Allan deviation of intensity fluctuations in a broadband frequency comb at three different wavelengths. At an averaging time of 10 minutes, typical for astronomical spectrographs, the uncertainty in intensity fluctuations is nearly an order of magnitude lower than at 10 seconds when real-time feedback is applied. In contrast, with a static SLM mask, intensity fluctuations exhibit a structured dependence on averaging time. These fluctuations, along with flux-dependent detector defects, remain challenging to correct in ongoing efforts to improve RV precision when no feedback is applied to maintain the intensity levels. 

%\section*{References} 

% Bibliography
\bibliography{supp_sample}

%Manual citation list
%\begin{thebibliography}{1}
%\bibitem{Zhang:14}
%Y.~Zhang, S.~Qiao, L.~Sun, Q.~W. Shi, W.~Huang, %L.~Li, and Z.~Yang,
 % \enquote{Photoinduced active terahertz metamaterials with nanostructured
  %vanadium dioxide film deposited by sol-gel method,} Opt. Express \textbf{22},
  %11070--11078 (2014).
%\end{thebibliography}